# Efficient Machine Learning Approach for Optimizing the Timing Resolution of a High Purity Germanium Detector

R. W. Gladen[1, *], V. A. Chirayath[1, ǂ], A. J. Fairchild[1], M. T. Manry[2], A. R. Koymen[1], and A. H. Weiss[1]

[1]*Department of Physics, University of Texas at Arlington, Arlington, TX, USA 76019*
[2]*Department of Electrical Engineering, University of Texas at Arlington, Arlington, TX, USA 76019*

[*]corresponding author e-mail: randall.gladen@uta.edu
[ǂ]corresponding author e-mail: chirayat@uta.edu

**Abstract**

We describe here an efficient machine-learning based approach for the optimization of parameters used for extracting the arrival time of waveforms, in particular those generated by the detection of 511 keV annihilation γ-rays by a 60 cm$^3$ coaxial high purity germanium detector (HPGe). The method utilizes a type of artificial neural network (ANN) called a self-organizing map (SOM) to cluster the HPGe waveforms based on the shape of their rising edges. The optimal timing parameters for HPGe waveforms belonging to a particular cluster are found by minimizing the time difference between the HPGe signal and a signal produced by a BaF$_2$ scintillation detector. Applying these variable timing parameters to the HPGe signals achieved a γ-γ coincidence timing resolution of ~ 4.3 ns at the 511 keV photo peak (defined as 511 $\pm$ 50 keV) and a timing resolution of ~ 6.5 ns for the entire γ spectrum—without rejecting any valid pulses. This timing resolution approaches the best obtained by analog nuclear electronics, without the corresponding complexities of analog optimization procedures. We further demonstrate the universality and efficacy of the machine learning approach by applying the method to the generation of secondary electron time-of-flight spectra following the implantation of energetic positrons on a sample.

**1. Introduction**

The time resolution of HPGe detectors are limited to several tens of nanosecond due to the large variance in the shape of the rising front of the pre-amplified HPGe signal and the electric noise [1-5]. The rising front of the signal reflects the point of interaction of the γ photon in the detector, the number of interactions per γ photon, as well as the charge transport dynamics across the electrodes, leading to an inverse relation between the volume of the HPGe detector/the energy of the γ photon and the time resolution [1-25]. Most of the methods—analog or digital—that aim to improve the timing resolution of the HPGe detector optimize the time pick-off algorithm by determining one set of time pick-off parameters for all the HPGe detector waveshapes. A single set of parameters, however, lead to large spread in time pick-off and often ~ 1 to 5 ns timing resolutions are attained only by rejecting a large fraction of waveshapes that are responsible for the time broadening [13-15]. The rejection of signals adversely affects the experimental count rate and can lead to loss of valuable information—especially for events involving low energy γ photons [5]. We demonstrate that it is possible to attain good timing resolution with minimal pulse rejection by utilizing time pick-off parameters that depend on the shape of the waveforms as described in detail later.

To obtain waveshape specific time pick-off parameters, we first cluster the pulses produced by a 60 $cm^3$ coaxial HPGe detector via an artificial neural network (ANN) with a self-organizing map (SOM) architecture [26]. This clustering is followed by a sub-procedure that minimizes the timing output variance of each cluster and is facilitated by the simultaneous detection of γ rays in an HPGe and a $BaF_2$ detector. The optimal parameters obtained through the minimization are generalized: given the same detector and similar γ source-detector geometry, the optimization procedure produces parameters that can be used for all future experiments. Implementing the optimal parameters yields a timing resolution of ~ 6.5 ns for the HPGe-$BaF_2$ detector combination placed inside a constant magnetic field without rejecting any valid pulses ("valid" meaning pulses that are due to true γ-ray detection and not false triggers) and including all γ energies from 40 keV up to 561 keV. Timing resolution is a complex function of the detector volume, detection geometry, γ energy, applied high voltage, electronic noise in the experimental system, the type of preamplifier employed, and the nature of algorithms employed. Here we have combined the clustering method with a modified form of the extrapolated leading-edge timing (ELET) algorithm to obtain the time information from the HPGe waveforms. The clustering method can, however, be combined with the digital implantation of other time pick-off techniques or noise filters [5, 13], or with shaping techniques such as moving window deconvolution [27, 28] to improve the timing and energy resolution achievable for a given semiconductor detection system.

Pioneering works that utilized analog processing methods to select specific pulse shapes had achieved a few ns (~ 1 ns) timing resolutions with small volume or planar Ge (Li) detectors [14, 15], and thus demonstrated the critical dependence of the timing resolution on waveform shape. Complex and time-consuming optimization of the timing filter amplifier (TFA)–constant fraction discriminator (CFD) combination was shown to obtain ~ 4 ns coincidence timing resolution with large volume HPGe detectors (~ 60 cm$^3$) that have ~ 10% efficiency [2] and that utilize γ photons only within 511±50 keV. However, such optimization methods are system specific and cannot be generalized to all semiconductor detector systems irrespective of the type, shape or size of the detectors or the energy of the γ photons detected. Subsequent attempts tried to address the significant loss in efficiency by either designing shape-compensated analog methods (achieving ~ 10 ns at 1.3 MeV with a 100 cm$^3$ HPGe) or resorting to offline time walk correction methods employing multiple ADCs with digital analysis (achieving ~ 3.4 ns at 511 keV photopeak with a 60 cm$^3$ HPGe) [16, 17].

The availability of fast digitizers have resulted in a slew of methods to increase the timing resolution of HPGe detectors, such as optimizing the parameters of digital time pick-off algorithms [5, 13] or implementing novel pulse shape discrimination techniques, which have the additional benefit of being able to determine the γ-ray interaction point in the detector for tracking or to discriminate between single site or multi-site interactions [3, 10, 12, 13, 18-25]. An example is the novel pulse shape analysis method by Crespi et al., [13] which performed a $\chi^2$ comparison of the input HPGe signal to a matrix of experimentally derived basis signals. The basis signals were selected to represent all possible time-shifted signal shapes that can be produced by the single site interaction of a γ-ray with the HPGe detector. This method yielded a coincident time resolution of 4.2 ns with a coaxial HPGe detector and included both the single site and multi-site interactions of 1.17 MeV γ photons with the detector. The SOM clustering and the optimization method we describe here, in comparison to all the digital methods listed above, is simple to implement and does not make any assumptions regarding the detector geometry or the nature of γ interactions and does not involve a complex $\chi^2$ minimization routine. The clustering of the pulse shapes was performed using the default SOM algorithm available in the Deep Learning Toolbox of MATLAB 2019b and similar implementations are widely available in all major Deep learning platforms. Clustering using SOM also has the advantage that it produces an artificial neural network, which, after being trained, efficiently classifies a new input waveform into the cluster with which it has maximum similarity. This makes the time pick-off process sufficiently fast using a trained network and optimized parameters. Machine learning techniques have been previously applied for the analysis of HPGe pulses, but they were only utilized to discriminate pulses due to single site and multisite interaction of γ photons [22-24]. To the best of our knowledge this is the first attempt that combines machine learning with timing algorithms.

Achieving nanosecond time resolution with HPGe detectors is critical in nuclear structure experiments using in-beam γ spectroscopy: the improved resolution can result in efficient background suppression and selection of events that originate at the target [5, 12, 13]. Pulse shape discrimination—like the clustering method described here—can aid in γ-ray tracking experiments with high-volume, segmented HPGe detectors [25, 28]. An improvement in the HPGe time resolution can also greatly enhance positron annihilation spectroscopies that involve timing with these detectors, such as the Age-Momentum Correlation (AMOC) technique or the measurement of the all three momentum components of the annihilating positron-electron pair [29-31].

Recently, our group has shown [32, 33], that timing using an HPGe detector with ~ 20 ns resolution—when employing a conventional digital ELET algorithm—can lead to a novel annihilation γ-$e^-$ coincidence technique that can provide unique surface electronic and chemical structure information. The γ-$e^-$ coincidence technique, first demonstrated by Eshed et al., and Kim et al., [34, 35] combines Positron annihilation induced Auger Electron Spectroscopy (PAES) [36] with the measurement of the Doppler broadening of the annihilation γ. The surface sensitivity of PAES [37, 38] in combination with Doppler Broadening Spectroscopy will provide a unique method to experimentally decompose the Doppler broadened spectra originating entirely from the annihilation of surface trapped positrons. The experimental decomposition of the Doppler broadened annihilation γ spectra can provide a valuable benchmark for theoretical calculations that aim to simulate the annihilation characteristics of positrons on surfaces [39]. Additionally, it may be the first step in standardizing the new inner surface characterization method entirely based on Doppler broadened γ spectroscopy [40, 41].

The Time-of-flight (ToF) PAES involves the measurement of the arrival time of the annihilation γ in coincidence with the arrival of the electron whereas the measurement of the Doppler broadening of the annihilation γ involves measuring the energy of the annihilation γ with a high energy resolution detector. A coincident measurement of all three necessary quantities (time of γ, energy of γ and time of $e^-$) if performed using three different detectors will require coincident detection of the two concurrently emitted antiparallel annihilation γ-rays in coincidence with the positron-induced electrons. This rigid condition would severely compromise the experiment's count rate and will preclude processes that involve the formation of ortho-Positronium (which annihilates predominantly via the emission of three γ photons). A three-parameter coincidence experiment also requires the employment of digital acquisition systems with more than two input channels. We have overcome all these drawbacks by consolidating the measurements of γ-ray timing and energy via a single HPGe detector. The method described in this paper has allowed us to determine the arrival time γ-rays with greater fidelity, thus improving the γ-$e^-$ coincidence technique. We provide a brief description of the experimental system in Sec. 2, followed by a detailed discussion of the computational methodologies adopted for the improvement of the HPGe time resolution in Sec. 3. The results of the technique along with its application in constructing the time-of-flight spectra of positron-induced secondary electrons are discussed in Sec. 4.

## 2. Experimental Apparatus

A detailed description of the apparatus is provided in [32] and an overview of the data analysis software can be found in [33]. A review is warranted here as the methods used to collect the data are relevant to the following discussions. The apparatus consists primarily of a variable energy positron beam, a sample chamber, and an electron time-of-flight spectrometer. In addition, the sample chamber is equipped with HPGe and BaF$_2$ detectors in an anti-parallel arrangement (Fig. 1), both of which are operated within a constant magnetic field of 50 Gauss. A mono-energetic beam of positrons (~17 eV) are incident upon a sample of zeolite that is biased to -500 V, which results in two processes: (1) the implantation of the positron in the material and its subsequent inelastic scattering will likely result in the emission of electrons into the vacuum (γ-$e^-$ coincidence), and (2) the eventual positron-electron annihilation creates two anti-parallel γ rays, each with an energy of ~ 511 keV, that are coincidentally detected by the HPGe and BaF$_2$ detectors (γ-γ coincidence). There are therefore three particles which may be collected in tandem, but as described above we consider only γ-$e^-$ coincidence and γ-γ coincidence in our experiments.

In γ-$e^-$ coincidence, the annihilation γ-ray is collected by the HPGe detector within a few hundred picoseconds of the initial positron implantation. The accompanying positron-induced electrons are guided to the electron detector (the "MCP" in Fig. 1) over a period varying from ~ 200 ns to 5 μs, depending on the initial electron energy. We then measure the flight time of these electrons as the time difference between the HPGe and electron detector pulses recorded by the oscilloscope.

In γ-γ coincidence the two simultaneously emitted γ-rays are collected by the HPGe and BaF$_2$ detectors. The HPGe detector used this experiment is an Ortec GEM 10195 (detecting volume of 60 cm$^3$) that lies along an axis additionally occupied by the corresponding BaF$_2$ detector. All the detectors mentioned transmit their pulses into a 12-bit Lecroy HDO 4200 oscilloscope with a sampling rate of 2.5 GS/s (Fig. 1).

## 3. The Clustering and the optimization procedure

We have illustrated the machine learning-based approach described in this paper as a schematic in Fig. 2. The primary steps are data collection (Fig. 2(a)), clustering and optimization (Fig. 2(b)), and application to γ-γ and γ-e$^-$ coincidence (Fig. 2 (c)). We have therefore collected four data sets: Datasets 1, 2 and 3 were collected using γ-γ coincidence and Dataset 4 with γ-e$^-$ coincidence. The clustering and optimization steps are applied only to the HPGe detector pulses; the time pickoff algorithms for the BaF$_2$ scintillation detector and the electron detector will be mentioned in the appropriate sections. The various steps in Fig. 2 are described in detail in sections 3.1, 3.2 and 3.3, respectively.

*3.1 Pulse Clustering*

To begin, we cluster the HPGe detector pulses (the input vectors) using the SOM algorithm provided by MATLAB® [42]. A variety of topologies can be selected for the SOM network from the Deep learning toolbox of MATLAB. We have selected an architecture consisting of a 32x32 neuron network with a hexagonal topology (Figure 2). This architecture was selected based on the size of the training data, the network training time, and the multitudinous shapes possible for the detector pulses. The network was trained for 440 iterations with a data set consisting of 255,000 HPGe detector pulses (Dataset 1 in Fig. 2). Prior to training, each pulse was corrected for baseline offset and cropped to a range of -50 μs and +250 μs (with respect to the oscilloscope trigger position) to focus on the rising edge of the pulse—i.e. the region of the pulse with the greatest shape variance. Fig. 3 exemplifies the pulse shapes prior to clustering, Fig. 4 represents the SOM network topology (i.e. the "sample hits") after clustering, and Fig. 5 demonstrates the results of clustering.

Fig. 4 is one of the many visualization tools provided by MATLAB for viewing the clustering results, in which the number of pulses associated with each neuron in the 32x32 hexagonal topology are given. After a sufficient number of training iterations, neighboring neurons in the SOM will be activated by detector pulses (or input vectors) with similar characteristics. We can clearly see this effect with Fig. 4, in which a large proportion of the input vectors are distributed into four distinct regions of the 2D space represented by the 32x32 network. Three of the regions are delimited by the purple, yellow, and green shapes in Fig. 4, along with a more diffuse, unmarked central region. The four clusters of pulses shown in Fig. 5—highlighted in red in Fig. 4—roughly typify these four different regions, with 50 representative pulses comprising each panel. The pulses in Fig 5(a) and 5(c) represent 511 keV γ photons (identified by their amplitude) that induced detector pulse shapes with different rise times, whereas Fig. 5(b) represents scattered annihilation γ; panel 5(d), however, represents false triggers due to low amplitude electronic noise or low amplitude photon-induced pulses. The efficacy of the clustering, as exemplified by Fig. 5, indicates that the SOM clustering algorithm may be used not only for timing optimization but also for understanding the interaction position of the γ-ray, as evidenced by its ability to identify shape variations on the rising edge of equal-amplitude pulses. Likewise, the clear demarcation of noise signals (Fig. 5(d)) automatically fulfills the requirements of a noise filter for rejecting false triggers during subsequent analyses.

We need to emphasize here that the data set used in the training of the clustering algorithm is never again used in the remaining optimization. To clarify, there are four altogether different data sets that are used throughout the procedure: (1) clustering, (2) parameter optimization with γ-γ coincidence, (3) implementation with γ-γ coincidence, and (4) implementation with γ-$e^-$ coincidence. Collecting these unique data sets for each step cross validates the clustering and avoids "over-training" the entire procedure to be compatible with only one particular set of data; the resulting parameters are thus generalized and applicable for all future experiments in our positron beam system for a given γ source-detector geometry.

*3.2 Initial Optimization*

We use in our timing analysis an algorithm based on extrapolated leading edge timing (ELET)—in which a straight line is fit to a region between two points (a "lower" value and an "upper" value) located at specific percentages (which are the time pick-off parameters in our method) of the pulse amplitude along the rising edge of the HPGe detector pulse [1]—but with a modification to the calculation of the pulse arrival time:

$$t_0 = \frac{-(f[1] + y_0)}{m}, \quad (1)$$

where $t_0$ is the extracted timing value, $f[1]$ is the first value in the linear fit vector (the lower percentage of the amplitude), $m$ is the slope, and $y_0$ is the y-intercept. Conventionally, with ELET, the time axis intersection of the extrapolated straight-line fit determines the pulse arrival time [28]. This works well when one uses time pickoff parameters (two points) which are very near the pulse baseline. However, an issue arises when fitting the straight line to a region well above the pulse baseline: small variations in the slope of the fit induce significant variations in the extrapolated crossover value (the very reason conventional ELET utilizes two points very close to the base line). To counter this effect, we find the horizontal intersection as defined by Eq. 1 instead.

The next step is to discover the two optimal amplitude percentages (i.e. the time pickoff parameters) through which the straight line is fit for each neural network cluster. We do this by creating multiple matrices (within a single "cell" structure in MATLAB) of the calculated time differences between the HPGe and $BaF_2$ detector pulses ($\Delta t$). Each matrix corresponds to a single neural network cluster—that is, only $\Delta t$ calculated using HPGe detector pulses that belong to a particular cluster are saved within each matrix, resulting in 1024 matrices corresponding to the 1024 neurons in the SOM network. Each column vector of each matrix represents a $\Delta t$ spectrum that was constructed with a particular pair of amplitude percentage values (lower and upper) to find the arrival time of the HPGe detector pulse, and the timing information from the coincident BaF2 pulses was calculated by applying Eq. 1 after a straight line fit between a single pair of amplitude percentage values (7%–9%). This was performed over a data set comprised of 50,000 HPGe/$BaF_2$ coincidence events (Dataset 2 in Fig. 2) and with ~ 200 amplitude percentage pairs. The lower of the two (p1) was varied from 1%–20% in steps of 2%, with the upper value varied through a range from p1 + 2% to p1 + 30% in steps of 4%.

Finally, histograms are created of the $\Delta t$ values for each amplitude percentage pair within each neuron cluster matrix. Each histogram is subsequently fit to a Gaussian to calculate the FWHM for every amplitude percentage pair: the pair with the minimum FWHM value is determined, as well as the centroid value of the corresponding Gaussian fit, which are then saved in yet another matrix. We have now ascertained the optimal parameters (the amplitude percentage pairs and the Gaussian centroid values) for each neural network cluster and can apply them to another unique data set. Section 3.1 and 3.2 is represented in Fig. 2 (b).

*3.3 Implementation*

Implementing the optimal parameters into the timing analysis of the new set of γ-γ or γ-$e^-$ coincidence is relatively straightforward: (i) the HPGe detector pulse in the γ-γ or γ-$e^-$ coincidence datum is first analyzed by the previously trained neural network and the cluster to which the pulse belongs is determined, (ii) the optimal amplitude percentages obtained for that cluster during the optimization procedure are inserted into the timing extraction algorithm and (iii) the extracted time from the HPGe detector pulse is shifted by the Gaussian centroid values assigned to the same cluster. The shift performed in the final step aligns the time spectrum of the HPGe detector pulses with the reference time obtained using the $BaF_2$ detector. Section 3.3 represents the steps shown in Fig. 2(c).

## 4. Results and Discussion

We obtained the γ-γ coincidence timing resolution reported here by analyzing Dataset 3 via the trained SOM network and by applying the optimal time pick off parameters for each pulse; likewise, we evaluated the applicability of the procedure for use in time-of-flight spectroscopy by using it to generate the time-of-flight spectra of positron-induced secondary electrons (Dataset 4). These results are discussed in detail below.

*4.1 γ-γ coincidence*

The histogram of the measured time differences (Δt) between the pulses produced by the $BaF_2$ detector and the HPGe detector is provided in Fig. 6. In this histogram we have included all detector pulses that were induced by a deposited γ energy of greater than 40 keV (including Compton scattering events), and we define the γ-γ coincidence timing resolution as the FWHM of the histogram obtained from a Lorentzian fit. The typical timing resolution of a $BaF_2$ detector is within the sub-nanosecond regime [43]; therefore the γ-γ coincidence timing resolution, which is a quadrature sum of the individual detector resolutions, is determined primarily by the timing resolution of the HPGe detector.

Fig. 6 (a) demonstrates the improvement in timing resolution of the HPGe detector achieved by the slight modification of the ELET algorithm as described by Eq. 1, with respect to the original algorithm given in Ref. [33]; but both of the histograms were constructed with the straight line fit on the rising edge of the pulse between 7% to 9% of the pulse amplitude. It's clear from the figure that the unmodified version of the algorithm results in a skewing of the peak towards the left (i.e. towards more negative times), likely due to pulses with relatively longer rise times. The timing resolution of the histogram constructed with the unmodified ELET algorithm is ~ 12.2 ns. We have found empirically that by redefining the ELET algorithm, as given by Eq. 1, we were able to reduce the skewness of the histogram while improving the resolution to ~ 10.1 ns.

The spread in the Δt histograms is due to the fact that a single pair of timing parameters (e.g. 7% to 9% of amplitude) cannot correct for all of the various line shapes, even when the parameters fit the straight line very near the baseline but above the baseline noise. In the timing optimization procedure, however, we select timing parameters optimized for each pulse shape—with different pulse shapes identified by SOM clustering. The result is a 35% improvement in timing resolution: a FWHM of 6.5 ns. To clarify, the discovery of the optimal timing parameters for each individual cluster reduces the time variance within that cluster, but applying different timing parameters for each cluster results in a large time variance between clusters. This effect is reduced by shifting the Δt histogram contribution of each individual cluster by the time-shift parameter (obtained via the Gaussian centroid) assigned to each cluster during the optimization step. This simply aligns all of the individual cluster contributions to the Δt histogram to the reference time as provided by the BaF2 detector. The application of this optimization procedure results in a symmetric Δt histogram that is effectively centered at zero. In literature, the coincidence timing resolution is nearly always reported for the γ photopeak of interest; and, to coincide with industry standards [2], we provide a Δt histogram (Fig. 6(c)) constructed by considering only those HPGe detector pulses that represent γ-ray energies within a window of $511\pm50$ keV. The timing resolution improves significantly to ~ 4.3 ns when selecting only the γ photopeak—comparable to the leading timing resolution reported so far at 511 keV using a large volume HPGe detector [2, 17].

*4.2 γ-e⁻ coincidence*

To assess the universality of the optimization procedure for our experimental system, we apply it to the analysis of Dataset 4, in which the HPGe detector pulses were collected in coincidence with positron-induced secondary electrons. We compare two Δt histograms of positron-induced secondary electrons: the first analyzed with the conventional ELET algorithm, and the second with the new machine learning-based ELET algorithm (Fig. 7). The higher Δt values on the horizontal axis correspond to low-energy electrons; and, likewise, the smaller Δt values correspond to higher energy electrons. The lower energy (higher Δt) edge in each peak corresponds to electrons that were emitted from the sample with zero energy and accelerated by the sample bias (-500 V in this case). The exponential decay apparent on the higher energy (lower Δt) part of each spectrum is due to the correlation of the detected electron with the incident positron that was reemitted from the sample as ortho-positronium, eventually decaying exponentially via three-gamma annihilation. This is reflected in the electron time-of-flight spectra as an exponential tail. The shapes of both spectra are consistent with what has been previously reported by our group [44]; and the spectrum constructed via the clustering and optimization procedure is distinguished by a much-improved shape at the higher Δt values, characterizing the improved timing resolution. This improvement is also evident by the clear separation of the exponentially decaying tail from the primary portion of the peak in the spectrum constructed using the clustering and optimization method.

*4.3 Applications of the machine learning approach*

We now present the benefits and some potential applications of the proposed procedure. There is first the simplicity of implementation: there is no need for a timing filter amplifier, digital or analog; there are only two parameters to optimize (the upper and lower amplitude percentages); and there are potentially a multitude of optimization algorithms that can be applied once the pulses have been clustered. The simplicity and versatility of the SOM clustering of the detector pulses endows any user with reasonable computing power the potential to greatly improve the resolution of their detectors—without the implementation of complex statistical analysis. For example, in our case, the clustering and the optimization routines were run on a single PC equipped with an Intel Xeon E3-1240 v5, 3.5 GHz processor with 64GB RAM. We would also like to emphasize that none of the algorithms were optimized to reduce the computational time and yet both training and optimization were completed in a reasonable time (less than a week). A secondary advantage is the range of γ energies that provide timing information with little loss in resolution—from ~ 40 keV to 560 keV.

The modified ELET algorithm and the described optimization procedure are not singular in their ability to discover the optimal timing parameters; for example, a method involving a one-parameter optimization (such as constant fraction discrimination) in tandem with a significantly larger clustering network may be used instead—as there would be no combinations of parameters to analyze, as there are in ELET—and the single parameter can be quickly optimized throughout a wide range of values for each cluster. The procedure described here represents only one route that may be taken towards the improvement of timing resolution in semiconductor detectors.

We additionally propose some alternate applications of the clustering method: As one example, the exact clustering and optimization procedure described here may be applied to the optimization of the timing resolution of a NaI detector with respect to a $BaF_2$ detector. This can potentially partially correct for the inverse scaling of the timing resolution with increasing NaI crystal volume and may have applications in ToF-PET [45]. As a second example, improvement in the energy resolution of an HPGe detector may be achieved by optimizing the parameters of a given shaping algorithm, such as moving window deconvolution [28], based on pulse shape. The clustering approach could also be applied to γ-ray tracking by identifying the interaction locations within the detector based on the pulse shape.

Finally, the potential for increasing the timing or energy resolution of particle detector pulses acquired with less expensive acquisition equipment was a primary motivator of this initial work; and we believe that—with the ever-increasing availability and simplicity of open-source machine learning packages—researchers with otherwise limited resources may be able to improve their results acquired with various particle detectors by applying such optimization procedures as those described here.

## 5. Conclusions

We have developed a new approach for improving the timing resolution of a semiconductor detector by finding the time pick-off parameters that work best for a particular shape of the detector output pulse. Various pulse shapes were identified and clustered using an ANN that was initially trained via unsupervised learning. The application of the trained network to the clustering of pulse shapes enabled the discovery of optimal time pick-off parameters, resulting in significant improvement in the timing resolution of a high volume (60 cm$^3$) coaxial HPGe detector. This approach has achieved, thus far, a γ-γ coincidence timing resolution of ~ 4.3 ns at 511 $\pm$ 50 keV (for an HPGe detector in coincidence with a BaF$_2$ detector) without the need for any complex data acquisition methods or complex statistical analysis. We further assessed the general applicability of the new timing method by applying it to the time-of-flight measurement of positron-induced secondary electrons.

**Acknowledgments**

We gratefully acknowledge the support by Welch Foundation (grant No. Y-1968-20180324) for the support during this work and for the development of the coincidence methods described in the manuscript. The positron beam system at UT Arlington was developed using the NSF major research instrumentation grant DMR-1338130 and NSF grant DMR-1508719.

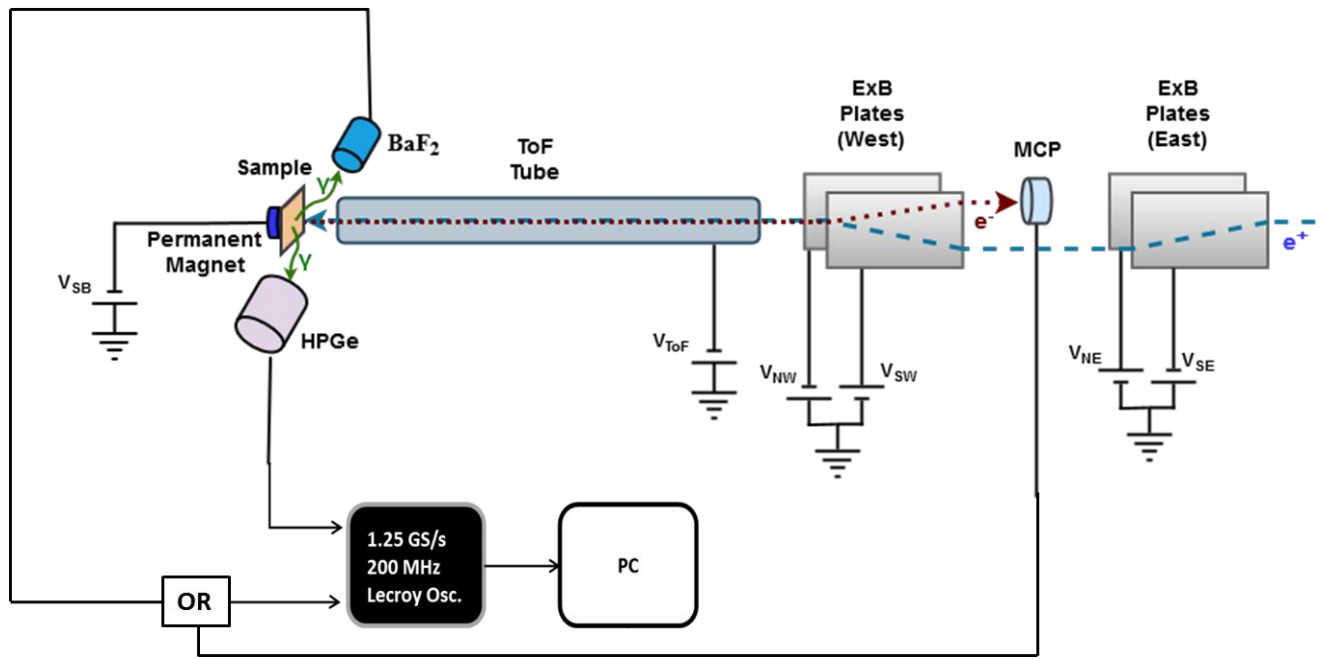

**Fig. 1** Schematic of the experimental setup used in the data collection. A beam of monoenergetic positrons (blue dashed line) is magnetically transported from the source to the sample. The positrons are bent around the electron detector (MCP) using a set of ExB plates. The positron annihilates with an electron of the sample, resulting in two antiparallel 511 keV annihilation γ-rays that are simultaneously detected by the BaF$_2$ and the HPGe detectors. Electrons (red dotted line) that are ejected from the sample as a result of positron implantation travel the length of the time-of-flight (ToF) tube until being drifted into the electron detector (MCP). The γ-γ coincidence setup proceeds by connecting the outputs of the BaF$_2$ and HPGe detectors to the oscilloscope. The BaF$_2$ connection is replaced by the MCP output for γ-$e^-$ coincidence measurements.

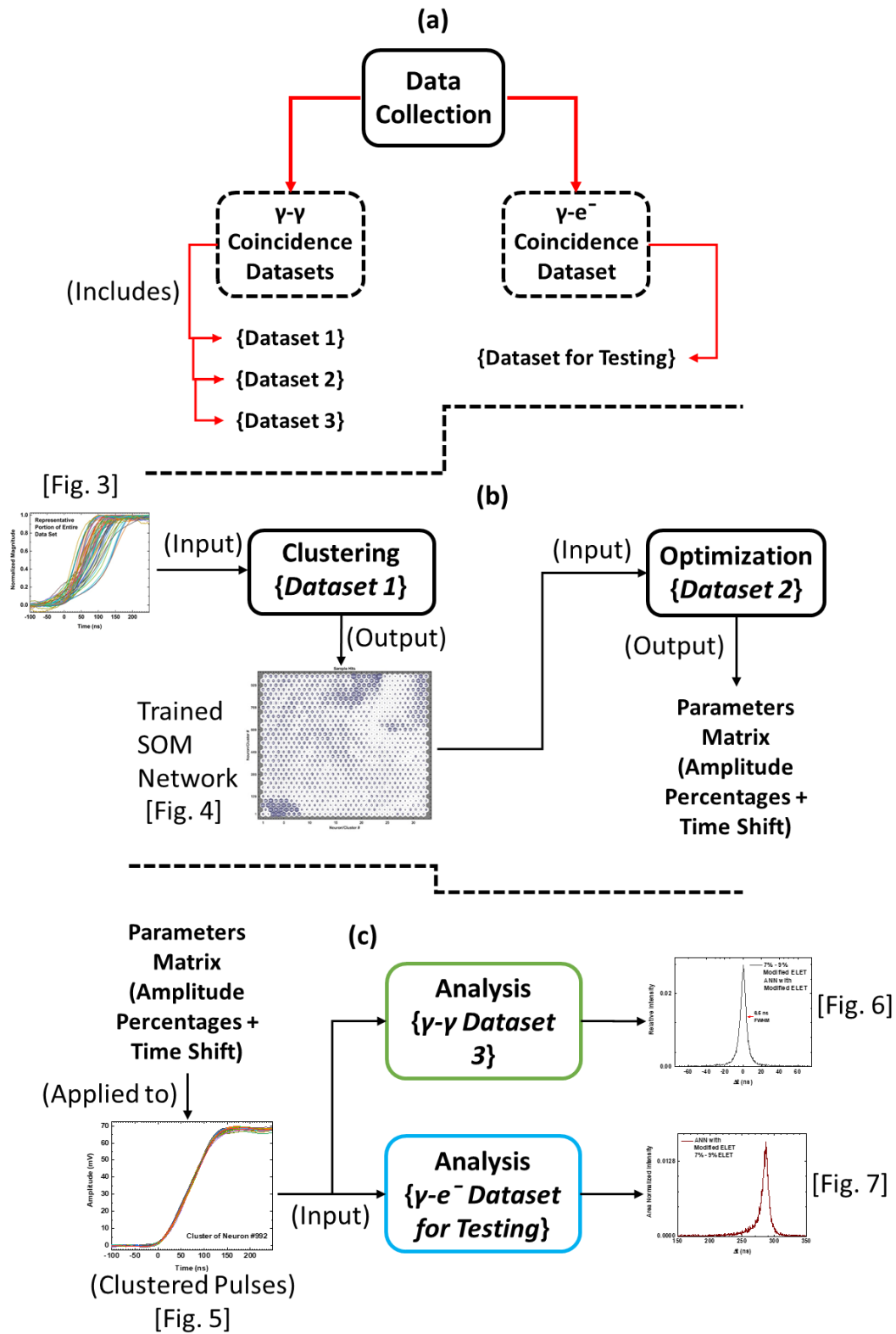

**Fig. 2** A series of flow diagrams to elucidate the workings of the clustering and optimization procedure. In panel (a), we begin by collecting four datasets: three γ-γ coincidence sets and one γ-e⁻ coincidence set. (b) The first dataset is used to train the SOM clustering network. The input vectors (as shown in Fig. 3) for training the SOM network are selected portions of the HPGe

output pulses of Dataset 1. The result is a trained SOM (as shown in Fig.4) that can then be used for identifying pulses in future datasets. This SOM then finds its application in the optimization procedure, in which the pulses from Dataset 2 are identified—an example of clustered pulses is provided in Fig. 5—using the SOM. Then, for each cluster, optimal time pick-off parameters and a time shift are found using the optimization routine. (c) The SOM and optimal timing parameters are applied to both Dataset 3 (in coincidence with the $BaF_2$ [γ] pulses) to obtain the coincidence timing resolution (see Fig. 6), and Dataset 4 (in coincidence with the MCP [$e^-$] pulses) to obtain the time-of-flight spectrum of positron-induced secondary electrons (see Fig. 7).

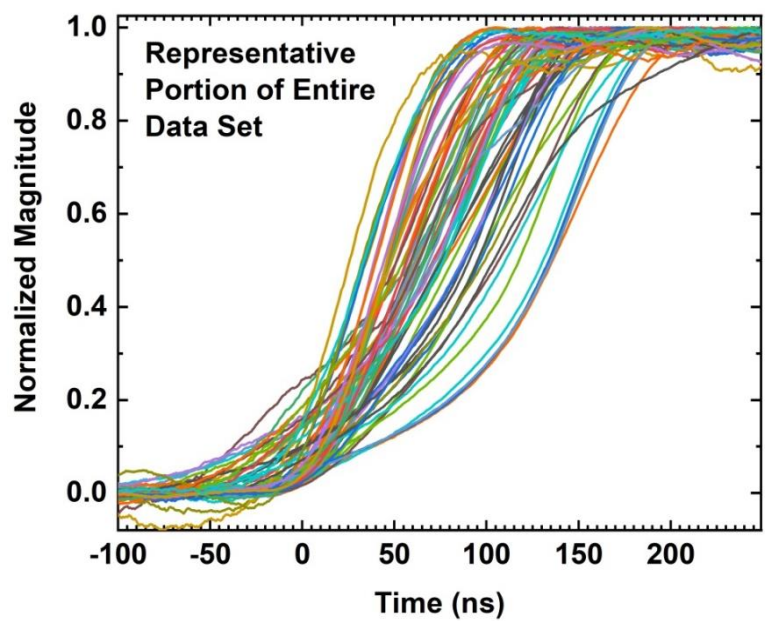

**Fig. 3** A representative portion of the HPGe detector pulses (67 pulses) that comprise the input vectors to the SOM neural network. The wide variation in the rise times of the output of the pre-amplified pulses can be seen here. The SOM network undergoes unsupervised training to cluster similar waveforms according to the specific characteristics of their individual shapes.

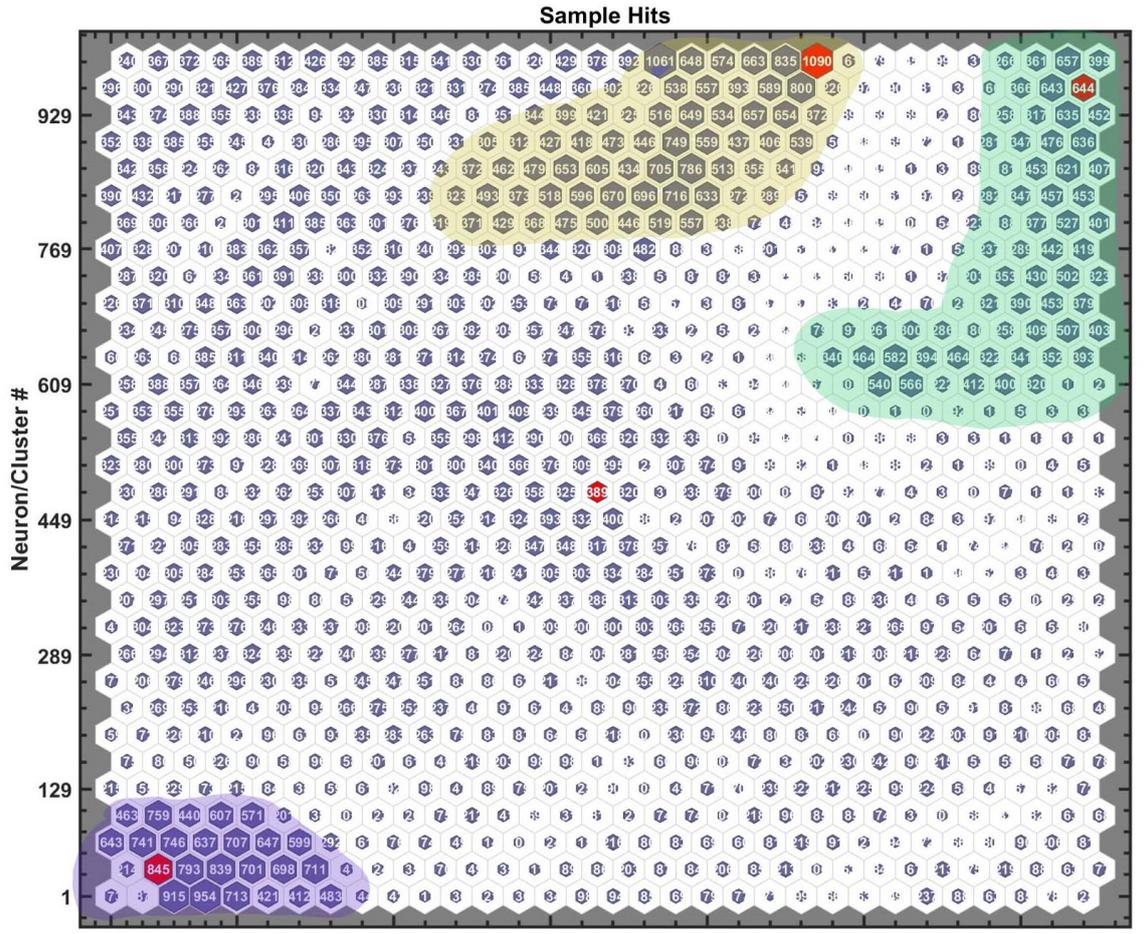

**Fig. 4** A "sample hits" plot of the 32x32 self-organizing map neural network. The sample hits represent a type of SOM visualization, in which the number of pulses associated with each neuron in the 32x32 hexagonal topology are given. The group of neurons highlighted in purple consist of pulses with relatively fast rise times, whereas the neurons highlighted in green consist of pulses with slower rise times. The yellow region consists primarily of noise and very low-amplitude pulses, and the rather diffuse, unmarked white region consists of pulses that represent inelastically scattered γ-rays within the detector. The red neurons were randomly selected to be roughly representative of each group of clusters, and their corresponding pulses are provided in Fig. 5.

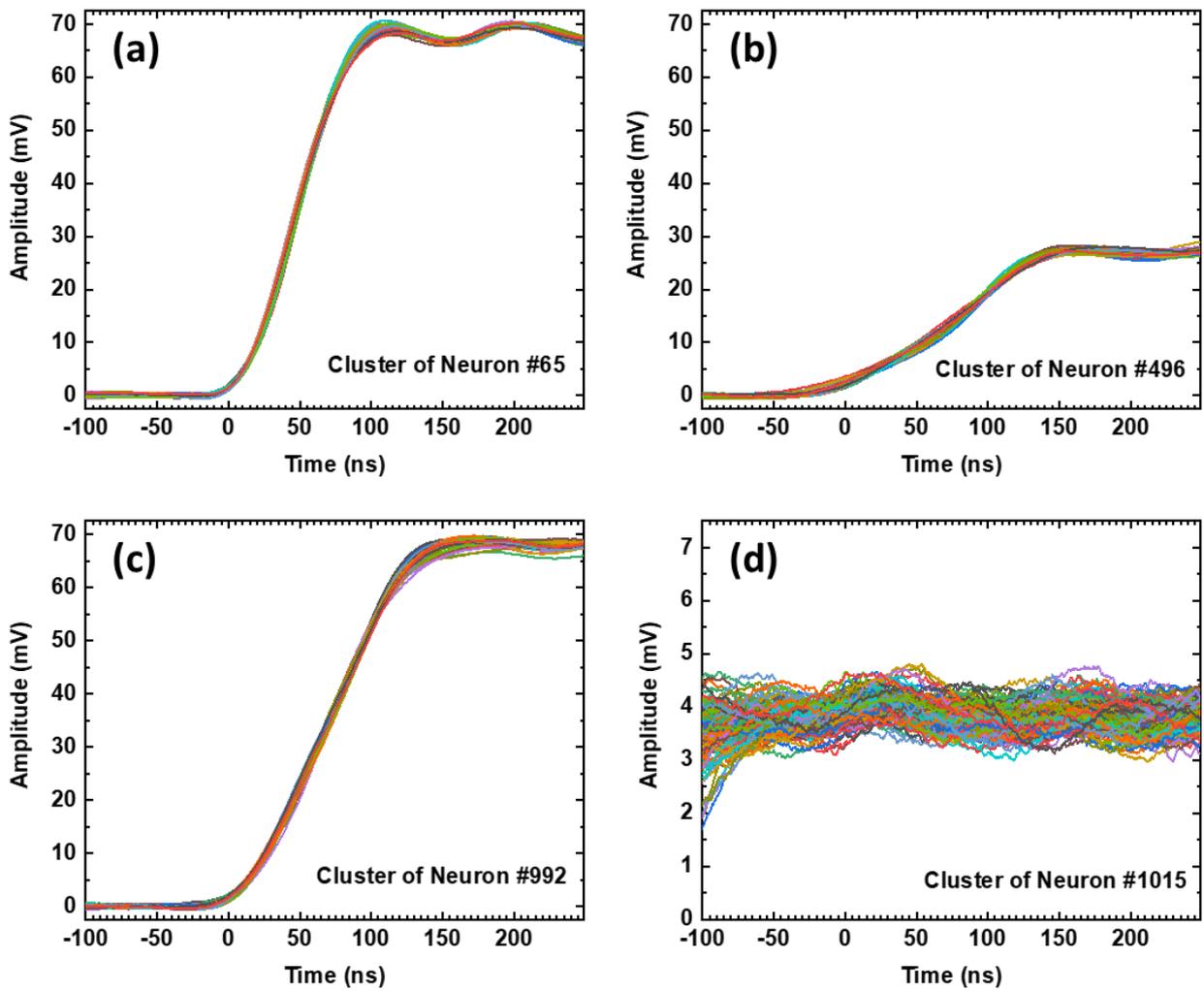

**Fig. 5** Examples of pulse clusters corresponding to four different neuron hits (50 pulses in each panel). (a) 511 keV annihilation γ-rays (identified by amplitude) that induced a relatively fast-rising pulse. (b) Detector pulses that were produced by the inelastic scattering of incident annihilation γ-rays in the detector. (c) 511 keV annihilation γ-rays that induced a relatively slow-rising pulse. (d) A cluster that corresponds to either false triggers or extremely low-amplitude pulses.

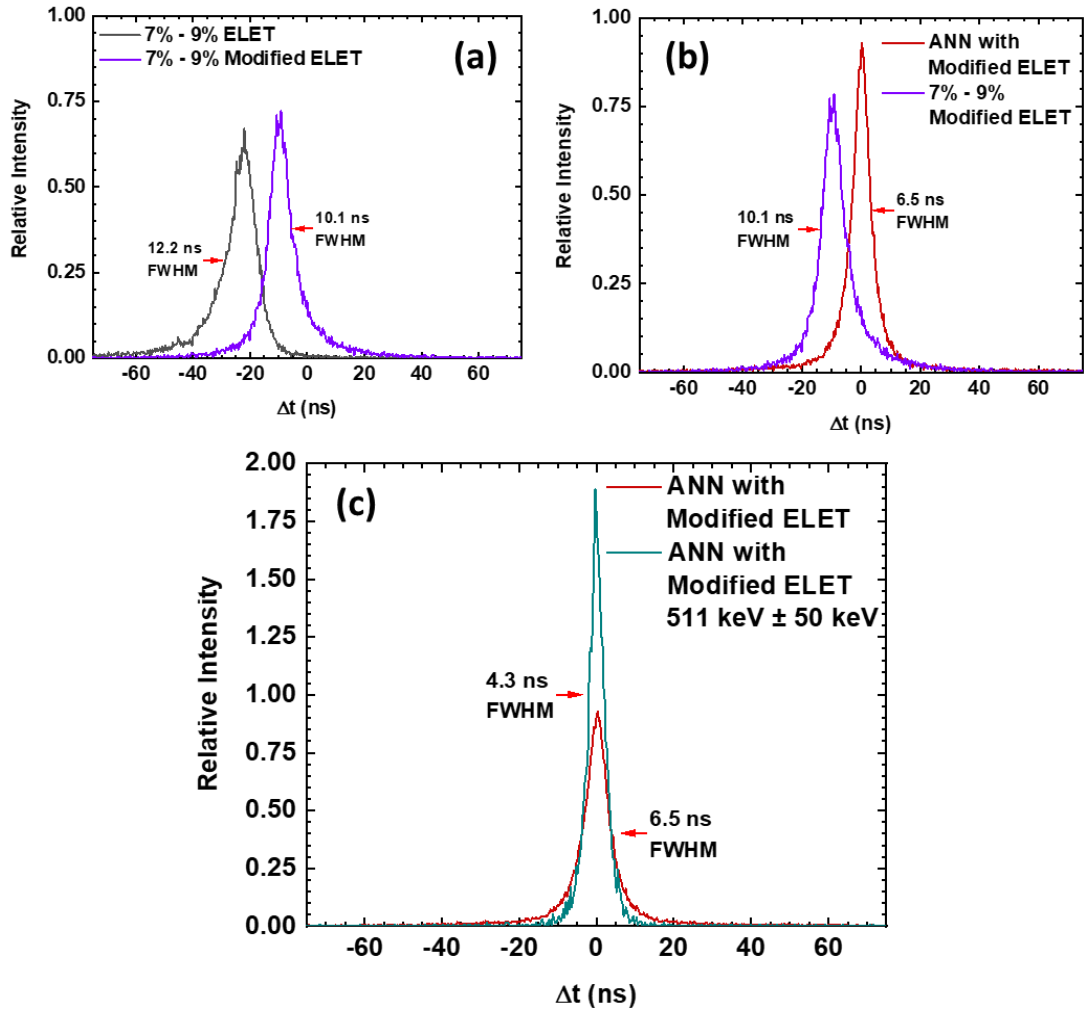

**Fig. 6** HPGe-BaF$_2$ coincidence peaks. For the creation of the γ-γ coincidence peaks all γ energies from 40 keV to 561 keV were utilized unless otherwise specified. The FWHM of the coincidence peaks is provided next to each peak, representing the coincidence timing resolution. (a) The grey peak was obtained by applying the conventional amplitude-based ELET algorithm to the HPGe and BaF$_2$ pulses. The larger FWHM and asymmetric distortion result from the variety of HPGe pulse shapes. In contrast, the violet peak was created with the modified ELET algorithm described in Section 3.2. (b) A comparison of the aforementioned peak constructed with the modified ELET algorithm (violet) and the peak resulting from this same modified ELET algorithm in tandem with the ANN clustering procedure (red). (c) The γ-γ coincidence spectra obtained by applying the modified ELET with ANN clustering (red) compared to the coincidence peak obtained with the same algorithm but considering only those γ-ray energies that lie within the 511 keV photopeak, defined as 511 $\pm$ 50 keV (cyan).

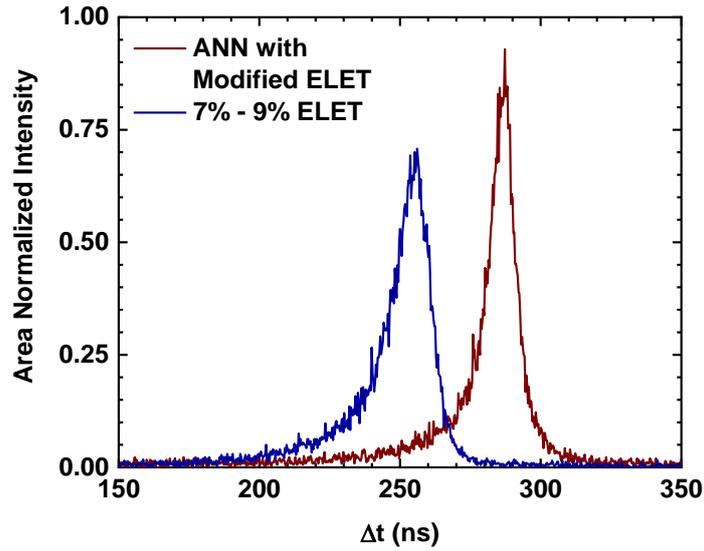

**Fig. 7** Time-of-flight spectra of the positron-induced secondary electrons. Both secondary electron spectra were constructed from the same data set (Data set 4) with the application of different timing methods as described in the legend. The histograms were constructed by measuring the time difference ($\Delta t$) between the detection of the annihilation γ-ray by the HPGe detector and the detection of the positron-induced electron by the electron detector. Here, a beam of monoenergetic positrons (~ 17 eV) is incident on a sample biased to -500 V (resulting in a positron energy of ~ 517 eV). The spectra generated using both methods are consistent with previously reported positron-induced secondary electron spectra [44]. The secondary electron spectrum generated with the ANN method distinguishes itself with important characteristics—such as a sharper low-energy (higher $\Delta t$) edge and a reduced FWHM—indicating improved timing resolution of the HPGe detector.